\documentclass[12pt]{article}
\usepackage{amsmath,amssymb,amsthm,amsxtra,overpic,bbm,bm,epsfig,subfigure}
\usepackage{mathrsfs}
\usepackage{graphicx}
\usepackage{color}
\usepackage{comment}
\usepackage{epstopdf}
\usepackage{float}
\usepackage{slashed,stmaryrd}

\def\thefootnote{\fnsymbol{footnote}}

\begin{document}

\vspace{0.2cm}

\begin{center}
{\Large\bf Matter effects on the flavor conversions of solar neutrinos and high-energy astrophysical neutrinos}
\end{center}

\vspace{0.2cm}

\begin{center}
{\bf Guo-yuan Huang} \footnote{E-mail: huanggy@ihep.ac.cn},
\quad
{\bf Jun-Hao Liu} \footnote{E-mail: liujunhao@ihep.ac.cn},
\quad
{\bf Shun Zhou} \footnote{E-mail: zhoush@ihep.ac.cn}
\\
{\small Institute of High Energy Physics, Chinese Academy of
Sciences, Beijing 100049, China \\
School of Physical Sciences, University of Chinese Academy of Sciences, Beijing 100049, China}
\end{center}

\vspace{1.5cm}

\begin{abstract}
Can we observe the solar eclipses in the neutrino light? In principle, this is possible by identifying the lunar matter effects on the flavor conversions of solar neutrinos when they traverse the Moon before reaching the detectors at the Earth. Unfortunately, we show that the lunar matter effects on the survival probability of solar $^8{\rm B}$ neutrinos are suppressed by an additional factor of $1.2\%$, compared to the day-night asymmetry. However, we point out that the matter effects on the flavor conversions of high-energy astrophysical neutrinos, when they propagate through the Sun, can be significant. Though the flavor composition of high-energy neutrinos can be remarkably modified, it is quite challenging to observe such effects even in the next-generation of neutrino telescopes.
\end{abstract}

\begin{flushleft}
\hspace{0.8cm} PACS number(s): 14.60.Pq, 13.15.+g, 26.65.+t
\end{flushleft}

\def\thefootnote{\arabic{footnote}}
\setcounter{footnote}{0}

\newpage

\section{Introduction}

The observations of solar neutrinos~\cite{McDonald:2016ixn} have led to the discovery of neutrino oscillations, together with those of atmospheric neutrinos~\cite{Kajita:2016cak}, revealing that neutrinos are massive and lepton flavors are significantly mixed~\cite{Patrignani:2016xqp}. The deficit of solar neutrinos $\nu^{}_e$ in the terrestrial detectors is now perfectly explained by neutrino flavor conversions under the Mikheyev-Smirnov-Wolfenstein (MSW) matter effects~\cite{Wolfenstein:1977ue, Mikheev:1986gs, Mikheev:1986wj}, together with a large mixing angle $\theta^{}_{12} \approx 34^\circ$ and a small neutrino mass-squared difference $\Delta m^2_{21} \equiv m^2_2 - m^2_1 \approx 7.5\times 10^{-5}~{\rm eV}^2$. Moreover, the enhanced neutrino flavor conversions caused by the Earth matter have been observed by comparing the neutrino events in the terrestrial detectors during the daytime with those during the nighttime. In the latter case, solar neutrinos have to pass through the Earth matter before entering into the detectors. The day-night asymmetry $A^{}_{\rm DN} \equiv 2(N^{}_{\rm D} - N^{}_{\rm N})/(N^{}_{\rm D} + N^{}_{\rm N}) = [-3.3 \pm 1.0 ~({\rm stat.}) \pm 0.5~(\rm syst.)]\%$ has been detected in the Super-Kamiokande experiment, where $N^{}_{\rm D}$ and $N^{}_{\rm N}$ stand for the numbers of solar neutrino events in the daytime and nighttime, respectively. This result is consistent with a nonzero asymmetry induced by the Earth matter at the $3\sigma$ level~\cite{Renshaw:2013dzu, Abe:2016nxk}.

Since solar neutrinos coming out of the Sun can be treated as the decoherent superposition of three neutrino mass eigenstates in vacuum, as was emphasized by Akhmedov in Ref.~\cite{Akhmedov:2000cs}, each mass eigenstate $|\nu^{}_i\rangle$ for $i = 1, 2, 3$ entering again into the medium of ordinary matter will finally induce significant flavor conversions even when they are propagating in matter for a relatively short distance. The day-night asymmetry of solar neutrinos serves as a typical example of this kind~\cite{Akhmedov:2000cs}. The effective Hamiltonian for neutrino flavor conversions in matter reads
\begin{eqnarray} \label{eq:Hm}
H^{}_{\rm m} = \frac{1}{2E} \left[ \left(\begin{matrix} m^2_1 & 0 & 0 \\ 0 & m^2_2 & 0 \\ 0 & 0 & m^2_3 \end{matrix}\right) + U^\dagger \left(\begin{matrix} A & 0 & 0 \\ 0 & 0 & 0 \\ 0 & 0 & 0 \end{matrix}\right)U \right] \; ,
\end{eqnarray}
after transforming from the flavor basis to the mass basis in vacuum, where $U$ stands for the flavor mixing matrix in vacuum, $E$ the neutrino energy and $m^{}_i$ for $i = 1, 2, 3$ neutrino mass eigenvalues. The matter term $A \equiv 2\sqrt{2} G^{}_{\rm F} N^{}_e E$ with $G^{}_{\rm F} = 1.166\times 10^{-5}~{\rm GeV}^{-2}$ being the Fermi constant and $N^{}_e$ the net electron number density, respectively, characterizes the contribution from the coherent forward scattering of the neutrinos propagating in a medium with the background particles. For the oscillations of antineutrinos, one can just perform the replacements $A \to - A$ and $U \to U^*$ in the effective Hamiltonian. Due to the second term in the square brackets in Eq.~(\ref{eq:Hm}), the transition between one neutrino mass eigenstate $|\nu^{}_i\rangle$ in vacuum to another one $|\nu^{}_j\rangle$ can be induced by the matter even if the quantum coherence among the initial mass eigenstates is completely lost. The observation of the day-night asymmetry of solar neutrinos in Super-Kamiokande demonstrates the correctness of this picture.

\begin{figure}[t]
\centering
\includegraphics[scale=0.8]{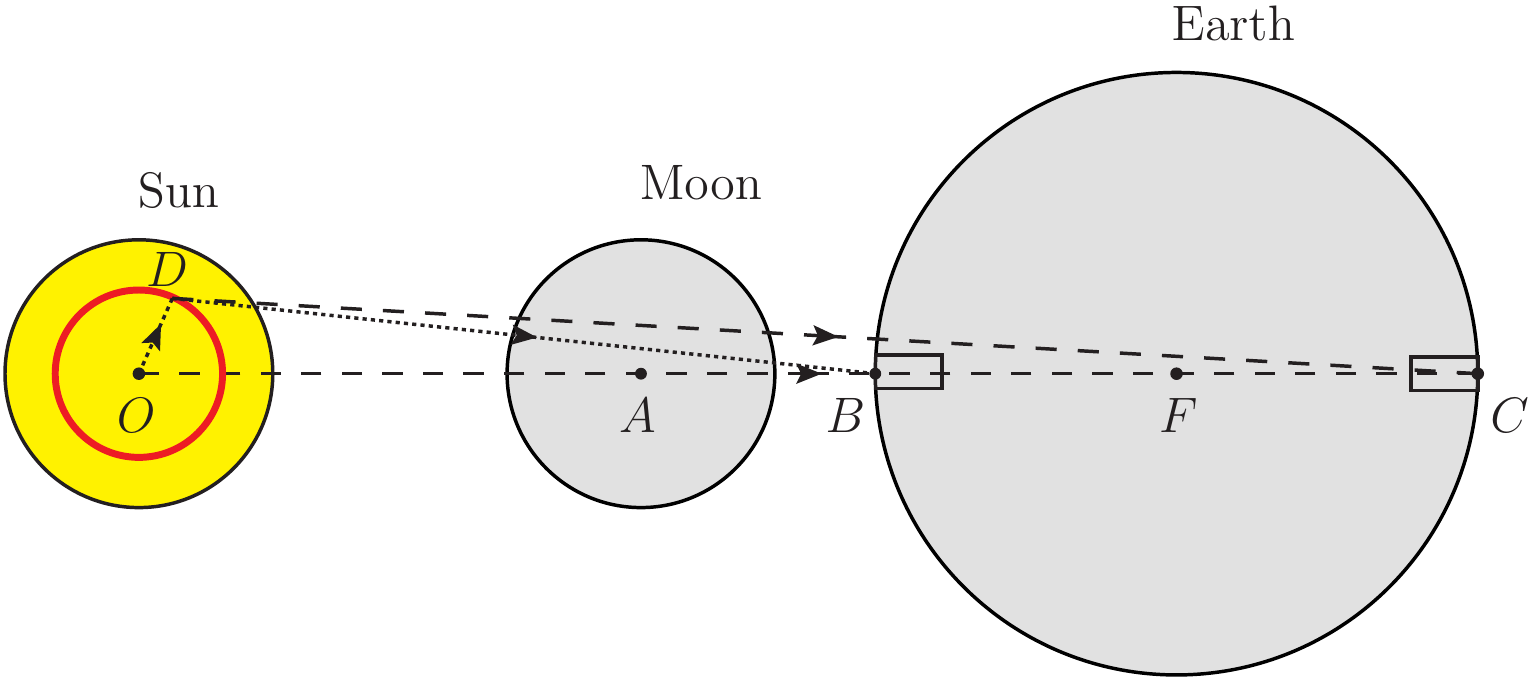}
\caption{The sketch for the positions of the Sun, the Moon and the Earth during a total solar eclipse. After solar neutrinos are emitted from the sphere of a radius $r = \overline{OD}$, they will traverse the Moon before arriving at the detector $B$ or at $C$ by further crossing the Earth.}
\label{fig:moon}
\end{figure}

In this work, we investigate whether it is possible to observe the solar eclipses in the neutrino light. During the solar eclipse, the Moon is located between the Sun and the Earth, so solar neutrinos have to pass through the Moon before arriving in the detector. See Fig.~\ref{fig:moon} for a brief explanation for the locations of the Sun, the Moon and the Earth when solar eclipses take place. Similar to the Earth matter effects, which are responsible for the day-night asymmetry, the lunar matter effects are expected to be of the same order. However, as we will show later, the distance between the Moon and the Earth is so long that the regenerated coherence between neutrino mass eigenstates emerging out of the Moon will be lost or averaged away, leaving a negligible impact on the survival probability of electron neutrinos from the Sun. In addition, we examine the lunar and solar matter effects on the high-energy astrophysical neutrinos, and demonstrate that the latter could be relevant for the high-statistics observations in future neutrino telescopes.

The remaining part of our work is structured as follows. In Section 2, we start with a special source of astrophysical neutrinos, which can be described as decoherent fluxes of neutrino mass eigenstates, and study how the lunar matter effects modify their survival probabilities. The matter effects on solar neutrinos and high-energy cosmic neutrinos are then discussed in some detail in Section 3. Finally, we summarize our main results in Section 4.

\section{Neutrino Flavor Conversions}
%%%%%%%%%%%%%%%%%%%%%%%%%%%%%%%%%%% Fig. 1 %%%%%%%%%%%%%%%%%%%%%%%%%%%%%%%%%
%%%%%%%%%%%%%%%%%%%%%%%%%%%%%%%%%%%%%%%%%%%%%%%%%%%%%%%%%%%%%%%%%%%%%%%%%%%%%
Without loss of generality, we consider the neutrino flavor eigenstate $|\nu^{}_\alpha\rangle$ for $\alpha = e, \mu, \tau$ from an astrophysical source, such as the solar neutrinos from the Sun and the high-energy neutrinos from Gamma Ray Bursts (GRBs) or Active Galactic Nuclei (AGN). Assuming $|\nu^{}_\alpha\rangle$ to be a decoherent superposition of three neutrino mass eigenstates $|\nu^{}_i\rangle$ for $i = 1, 2, 3$, we can find out the transitional probabilities $P^{}_{\alpha \beta} \equiv P(\nu^{}_\alpha \to \nu^{}_\beta)$ if there are no other media along the way to the detector
\begin{eqnarray} \label{eq:PS}
P^{}_{\alpha \beta} = \sum^3_{i=1} k^{\alpha}_i |U^{}_{\beta i}|^2 \; ,
\end{eqnarray}
where $U^{}_{\beta i}$ for $\beta = e, \mu, \tau$ and $i = 1, 2, 3$ are the elements of the leptonic flavor mixing matrix $U$ in vacuum, $k^{\alpha}_i$ denotes the fraction of $|\nu^{}_i\rangle$ contained in the initial flavor state $|\nu^{}_\alpha\rangle$. Note that the normalization condition $k^{\alpha}_1 + k^{\alpha}_2 + k^{\alpha}_3 = 1$ is satisfied for each individual neutrino flavor. The results in Eq.~(\ref{eq:PS}) can be understood as follows: each mass eigenstate $|\nu^{}_i\rangle$ arrives in the detector and will be projected to the flavor eigenstate $|\nu^{}_\beta\rangle$ with a probability of $|U^{}_{\beta i}|^2$.

If there is an astrophysical object, such as the Moon, standing in the way between the neutrino source and the detector, the mass eigenstate $|\nu^{}_i\rangle$ in vacuum will enter into the medium and then exit it with a probability to be another mass eigenstate $|\nu^\prime_j\rangle$. It should be noticed that the prime in $|\nu^\prime_j\rangle$ is just used to discriminate between the neutrino mass eigenstates in vacuum before and after traversing the medium. For simplicity, we assume that the distance traveled by neutrinos inside the astrophysical object is just its diameter $d^{}_{\rm M}$ and the matter density $\rho^{}_{\rm M}$ is constant with an electron number fraction $Y^e_{\rm M}$. In addition, neutrinos emerging out of this object will propagate for a distance $L$ to reach the detector, which is supposed to be sensitive the neutrino flavor state $|\nu^{}_\beta\rangle$. Before computing the transitional probability $\widehat{P}^{}_{\alpha \beta} = P(\nu^{}_\alpha \to \nu^{}_\beta)$ in this case, we have to distinguish two different scenarios:
\begin{itemize}
\item The distance $L$ happens to be so long that the overlap among the neutrino mass eigenstates $|\nu^\prime_j\rangle$ for $j = 1, 2, 3$ in vacuum disappears before they enter into the detector. Therefore, the final transitional probabilities are given by
\begin{eqnarray}\label{eq:PS1}
\widehat{P}^{\rm dec}_{\alpha \beta} = \sum^{3}_{i = 1} \sum^{3}_{j = 1} k^\alpha_i P(\nu^{}_i \to \nu^\prime_j) |U^{}_{\beta j}|^2 \; ,
%     (3)
\end{eqnarray}
where $P(\nu^{}_i \to \nu^\prime_j) \equiv P^{}_{ij}$ stands for the transitional probability for $|\nu^{}_i\rangle \to |\nu^\prime_j\rangle$ after passing through the medium.

\item Different from the previous scenario, the coherence could be maintained when the distance $L$ is comparable to the coherent length, which can only be estimated after specifying the sizes of wave packets of neutrinos in production. Without the details of neutrino production, we just compare between the oscillation length $L^{ij}_{\rm osc} \equiv 4\pi E/|\Delta m^2_{ij}|$, where $\Delta m^2_{ij} \equiv m^2_i - m^2_j$ is the relevant neutrino mass-squared difference in question, and the traveling distance. For clarity, we assume that the coherence is retained and $L^{ij}_{\rm osc}$ is comparable to $L$, so the final transitional probabilities can be written as
\begin{eqnarray}\label{eq:PS2}
\widehat{P}^{\rm coh}_{\alpha \beta} = \sum^3_{i = 1} k^\alpha_i P(\nu^{}_i \to \nu^{}_\beta) \; ,
%     (4)
\end{eqnarray}
where $P(\nu^{}_i \to \nu^{}_\beta) \equiv P^{}_{i\beta}$ are the transitional probabilities for $|\nu^{}_i\rangle \to |\nu^{}_\beta\rangle$, in which the neutrino mass eigenstates $|\nu^\prime_j\rangle$ appear as the intermediate states.
\end{itemize}
Since the decoherent scenario can be treated as a special case of the coherent scenario when the interference terms are averaged out, we proceed with the calculation of $\widehat{P}^{\rm coh}_{\alpha \beta}$ in Eq.~(\ref{eq:PS2}) and derive $\widehat{P}^{\rm dec}_{\alpha \beta}$ by removing the interference terms. Explicitly, we have
\begin{eqnarray}\label{eq:Pab}
\widehat{P}^{\rm coh}_{\alpha \beta} = \sum^3_{i = 1} \sum^3_{j = 1} k^\alpha_i \left|\langle \nu^{}_\beta|\nu^\prime_j\rangle \cdot \exp\left[-{\rm i} m^2_j L/(2E)\right] \cdot \langle \nu^\prime_j |\nu^{}_i\rangle \right|^2 = \widehat{P}^{\rm dec}_{\alpha \beta} + \widehat{I}^{}_{\alpha \beta} \; ,
%     (5)
\end{eqnarray}
and the interference terms are given by
\begin{eqnarray}\label{eq:Iab}
\widehat{I}^{}_{\alpha \beta} = \sum^3_{i = 1} k^\alpha_i \sum_{j > k} 2 {\rm Re}\left\{ U^{}_{\beta j} U^*_{\beta k} A^{}_{ij} A^*_{ik} \exp\left[-{\rm i}\Delta m^2_{jk} L/(2E)\right]\right\} \; ,
%     (6)
\end{eqnarray}
where $\Delta m^2_{jk} \equiv m^2_j - m^2_k$ for $jk = 21, 31, 32$ are neutrino mass-squared differences, $A^{}_{ij} \equiv \langle \nu^\prime_j|\nu^{}_i\rangle$ denotes the transitional amplitude for $|\nu^{}_i\rangle \to |\nu^\prime_j\rangle$ and the corresponding probability is $P^{}_{ij} \equiv P(\nu^{}_i \to \nu^\prime_j) = |A^{}_{ij}|^2$. Now it is clear that we have to calculate the transitional amplitude for a neutrino mass eigenstate $|\nu^{}_i\rangle$ to exit the medium as $|\nu^\prime_j\rangle$.

As we have assumed that the matter density $\rho^{}_{\rm M}$ of the astrophysical object is constant, the matter term in Eq.~(\ref{eq:Hm}) is given by $A = 2\sqrt{2}G^{}_{\rm F} N^{}_e E$ with $N^{}_e = Y^e_{\rm M} N^{}_{\rm A} [\rho^{}_{\rm M}/(1~{\rm g}~{\rm cm}^{-3})]~{\rm cm}^{-3}$ with $N^{}_{\rm A} = 6.022\times 10^{23}$ being the Avogadro constant. It is straightforward to diagonalize the effective Hamiltonian by the unitary matrix $V$ via
\begin{eqnarray}\label{eq:diag}
V^\dagger H^{}_{\rm m} V = {\rm diag}\{\tilde{m}^2_1, \tilde{m}^2_2, \tilde{m}^2_3\}/(2E) \; ,
%     (7)
\end{eqnarray}
where $\tilde{m}^{}_i$ (for $i = 1, 2, 3$) stand for the effective neutrino masses in matter. It is worth mentioning that the corresponding effective mixing matrix is $\tilde{U} = U V$, as the transformation from the flavor basis to the vacuum mass basis has been performed in Eq.~(\ref{eq:Hm}). The explicit expressions of effective neutrino mass eigenvalues can be found in Ref.~\cite{Xing:2000gg} and are quoted below
\begin{eqnarray}\label{eq:mtilde}
\tilde{m}^2_1 &=& m^2_1 + \frac{1}{3}x - \frac{1}{3} \sqrt{x^2 - 3y} \left[z + \sqrt{3(1 - z^2)}\right] \; , \nonumber \\
\tilde{m}^2_2 &=& m^2_1 + \frac{1}{3}x - \frac{1}{3} \sqrt{x^2 - 3y} \left[z - \sqrt{3(1 - z^2)}\right] \; , \nonumber \\
\tilde{m}^2_3 &=& m^2_1 + \frac{1}{3}x + \frac{2}{3} z \sqrt{x^2 - 3y} \; ,
%     (8)
\end{eqnarray}
where
\begin{eqnarray}\label{eq:xyz}
x &=& \Delta m^2_{21} + \Delta m^2_{31} + A \; , \nonumber \\
y &=& \Delta m^2_{21} \Delta m^2_{31} + A \left[\Delta m^2_{21} (1 - |U^{}_{e2}|^2) + \Delta m^2_{31}(1 - |U^{}_{e3}|^2)\right] \; , \nonumber \\
z &=& \cos \left[\frac{1}{3} \arccos \frac{2x^3 - 9xy + 27 A \Delta m^2_{21} \Delta m^2_{31} |U^{}_{e1}|^2}{2(x^2 - 3y)^{3/2}}\right] \; ,
%     (9)
\end{eqnarray}
and the normal neutrino mass ordering (i.e., $\Delta m^2_{31} > 0$) is assumed. Moreover, the matrix elements of $V$ have also been derived in Ref.~\cite{Xing:2000gg}:
\begin{eqnarray}\label{eq:Vij}
V^{}_{ii} = \frac{N^{}_i}{D^{}_i} \; , \quad V^{}_{ij} = \frac{A}{D^{}_j} \left(\tilde{m}^2_j - m^2_k\right) U^*_{ei} U^{}_{ej} \; ,
%     (10)
\end{eqnarray}
where $i, j, k$ run over $1, 2, 3$ with $i \neq j \neq k$, and
\begin{eqnarray}\label{eq:nd}
N^{}_i &=& (\tilde{m}^2_i - m^2_j)(\tilde{m}^2_i - m^2_k) - A \left[ (\tilde{m}^2_i - m^2_j) |U^{}_{ek}|^2 + (\tilde{m}^2_i - m^2_k) |U^{}_{ej}|^2\right] \; , \nonumber \\
D^2_i &=& N^2_i + A^2 |U^{}_{ei}|^2 \left[ (\tilde{m}^2_i - m^2_j)^2 |U^{}_{ek}|^2 + (\tilde{m}^2_i - m^2_k)^2 |U^{}_{ej}|^2\right] \; .
%     (11)
\end{eqnarray}
We stress that $V$ in our discussions is not just an intermediate step to derive the mixing matrix $\tilde{U}$ as in Ref.~\cite{Xing:2000gg}, but useful to calculate the transitional amplitudes $A^{}_{ij}$ or the probabilities $P^{}_{ij}$. With the help of Eqs.~(\ref{eq:mtilde}) and (\ref{eq:Vij}), we immediately obtain the transitional amplitudes and probabilities for $|\nu^{}_i\rangle \to |\nu^\prime_j\rangle$ by following the evolution of neutrino mass eigenstates $|\tilde{\nu}^{}_k\rangle$ in matter
\begin{eqnarray}
\label{eq:Aij}
A^{}_{ij} &=& \langle \nu^\prime_j|\nu^{}_i\rangle = \sum^3_{k = 1} V^{}_{jk} V^*_{ik} \exp\left[-{\rm i}\tilde{m}^2_k d^{}_{\rm M}/(2E)\right] \; , \\
%     (12)
\label{eq:Pij}
P^{}_{ij} &=& \sum^3_{k=1} |V^{}_{ik}|^2 |V^{}_{jk}|^2 + \sum^{}_{m > n} 2{\rm Re}\left\{V^{}_{in} V^{}_{jm} V^*_{im} V^*_{jn} \exp\left[-{\rm i} \Delta \tilde{m}^2_{mn} d^{}_{\rm M}/(2E)\right]\right\} \; ,
%     (13)
\end{eqnarray}
where $\Delta \tilde{m}^2_{mn} \equiv \tilde{m}^2_m - \tilde{m}^2_n$ for $mn = 21, 31, 32$ are the neutrino mass-squared differences in matter. Hence the transitional probabilities in Eqs.~(\ref{eq:PS1}) and (\ref{eq:PS2}) can readily be calculated by using Eqs.~(\ref{eq:Aij}) and (\ref{eq:Pij}). If neutrinos propagate through the Earth before arriving at the detector, the Earth matter effects can also be further taken into account in a similar way.

\section{Lunar and Solar Matter Effects}

\subsection{Solar Neutrinos}
First, let us apply the formalism in the previous section to solar neutrinos. The matter effects on the flavor conversions of solar neutrinos inside the Sun and the Earth have been extensively studied in the literature. See, e.g., Refs.~\cite{Blennow:2013rca, Maltoni:2015kca}, for recent reviews on this topic. The electron neutrino state $|\nu^{}_e\rangle$ coming out of the Sun can be expressed as the decoherent superposition of three neutrino mass eigenstates $|\nu^{}_i\rangle$ (for $i = 1, 2, 3$). For solar neutrinos, the fraction of the mass eigenstate $|\nu^{}_i\rangle$ in $|\nu^{}_e\rangle$ at the surface of the Sun is given by~\cite{Blennow:2003xw}
\begin{eqnarray} \label{eq:ki}
k^e_i = \sum^3_{j=1}\int^{R^{}_{\rm S}}_0 {\rm d}r f(r) |\tilde{U}^{}_{ej}(r)|^2 P^{\rm m}_{ji} \; ,
%     (14)
\end{eqnarray}
where $f(r)$ is the normalized distribution function of solar neutrinos, characterizing the fraction of neutrino production at a distance $r$ in the solar core, and $R^{}_{\rm S}$ is the solar radius. In addition, we have introduced the probability $P^{\rm m}_{ij} \equiv P(\tilde{\nu}^{}_i \to \nu^{}_j)$ for solar neutrinos that are produced in the core as a mass eigenstate $|\tilde{\nu}^{}_i\rangle$ to be a mass eigenstate $|\nu^{}_j\rangle$ at the surface. As the change of solar matter density is sufficiently slow, an adiabatic evolution of neutrino mass eigenstates is guaranteed and thus $P^{\rm m}_{ij} \equiv P(\tilde{\nu}^{}_i \to \nu^{}_j)$ is essentially vanishing for $i\neq j$. The exact values of $k^e_i$ (for $i = 1, 2, 3$) can be found in Ref.~\cite{Blennow:2003xw} and some references therein.

To completely study the matter effects, we need to calculate the survival probability of solar neutrinos in the daytime $P^{}_{\rm S}$ and those in another three different cases: (A) Neutrinos pass through only the Earth; (B) Neutrinos traverse the Moon but not the Earth; (C) Neutrinos go through both the Moon and the Earth. The survival probabilities in these three cases will be denoted as $P^{}_{\rm SE}$, $P^{\rm M}_{\rm S}$ and $P^{\rm M}_{\rm SE}$, respectively. For the ordinary day-night effects, the relevant quantity is the difference between $P^{}_{\rm S}$ and $P^{}_{\rm SE}$. In the framework of three-flavor neutrino mixing, for which the mixing matrix $U$ is conventionally parametrized in terms of three mixing angles $\{\theta^{}_{12}, \theta^{}_{13}, \theta^{}_{23}\}$ and one CP-violating phase $\delta$~\cite{Patrignani:2016xqp}, it has been found~\cite{Blennow:2003xw}
\begin{eqnarray} \label{eq:pseps}
P^{}_{\rm SE} - P^{}_{\rm S} = - \cos^6\theta^{}_{13} \frac{\Delta m^2_{21} A}{\left(\Delta \tilde{m}^2_{21}\right)^2}\sin^2 2\theta^{}_{12} \sin^2 \left(\frac{\Delta \tilde{m}^2_{21} L^{}_{\rm E}}{4E}\right)\int^{R^{}_{\rm S}}_0 {\rm d}r f(r) \cos 2\tilde{\theta}^{}_{12}(r) \; ,
%     (15)
\end{eqnarray}
where $\Delta \tilde{m}^2_{21} \equiv [(\Delta m^2_{21} - A\cos^2\theta^{}_{13} \cos 2\theta^{}_{12})^2 + A^2 \cos^4\theta^{}_{13} \sin^2 2\theta^{}_{12}]^{1/2}$ is the effective neutrino mass-squared difference in matter, $L^{}_{\rm E}$ is the distance that neutrinos have traveled in the Earth, and $\tilde{\theta}^{}_{12}(r)$ is the effective mixing angle in matter. For the $^8{\rm B}$ neutrinos produced in the solar core, the matter density is sufficiently large so that $\tilde{\theta}^{}_{12}(r)$ is close to $\pi/2$, indicating $\cos 2\tilde{\theta}^{}_{12}(r) < 0$ and $P^{}_{\rm SE} > P^{}_{\rm S}$, which is well consistent with the observation of $A^{}_{\rm DN} < 0$ in the Super-Kamiokande experiment, as mentioned in the introduction.

The average distance between the Moon and the Earth is $L^{}_{\rm ME} \approx 3.84\times 10^5~{\rm km}$, which is much larger than the oscillation length of solar $^8{\rm B}$ neutrinos, namely,
\begin{eqnarray}\label{eq:Losc}
L^{}_{\rm osc} \sim L^{21}_{\rm osc} \equiv \frac{4\pi E}{\Delta m^2_{21}} \approx 330~{\rm km}~\left(\frac{E}{10~{\rm MeV}}\right) \cdot \left(\frac{7.5\times 10^{-5}~{\rm eV}^2}{\Delta m^2_{21}}\right) \; ,
%     (16)
\end{eqnarray}
so it is reasonable to assume that the interference terms in Eq.~(\ref{eq:Iab}) should be averaged away. As a consequence, the survival probability including the lunar matter effects turns out to be
\begin{eqnarray}\label{eq:PMS}
P^{\rm M}_{\rm S} = \sum^{3}_{i = 1} \sum^3_{j = 1} k^e_i P^{}_{ij} |U^{}_{ej}|^2 \; ,
%     (17)
\end{eqnarray}
which can be further simplified in light of neutrino oscillation data. First, due to the smallness of $|U^{}_{e3}|^2 = \sin^2 \theta^{}_{13} \approx 0.02$, the summation over the index $j$ in Eq.~(\ref{eq:PMS}) can be reduced to the first two mass eigenstates $|\nu^\prime_1\rangle$ and $|\nu^\prime_2\rangle$. Second, since the mass-squared difference $\Delta m^2_{31} \approx 2.5\times 10^{-3}~{\rm eV}^2$ is much larger than the matter term $A \approx 1.52\times 10^{-6}~{\rm eV}^2~Y^{}_e~[\rho^{}_{\rm c}/(1~{\rm g}~{\rm cm}^{-3})]\cdot[E/(10~{\rm MeV})]$ in the solar core with $Y^{}_e \approx 0.67$ and $\rho^{}_{\rm c} \approx 150~{\rm g}~{\rm cm}^{-3}$, we have $|\tilde{U}^{}_{e3}|^2 \approx |U^{}_{e3}|^2$, leading to $k^e_3 \approx |U^{}_{e3}|^2 \ll 1$. Therefore, if the higher-order terms ${\cal O}(\sin^2 \theta^{}_{13})$ are neglected, we obtain
\begin{eqnarray}\label{eq:approx}
P^{\rm M}_{\rm S} - P^{}_{\rm S} = (k^e_2 - k^e_1) (|U^{}_{e1}|^2 - |U^{}_{e2}|^2) P^{}_{12} \approx - P^{}_{12} \cos^4 \theta^{}_{13} \cos 2\theta^{}_{12} \int^{R^{}_{\rm S}}_0 {\rm d}r f(r) \cos 2\tilde{\theta}^{}_{12}(r) \; ,
%     (18)
\end{eqnarray}
where the transitional probability $P^{}_{12}$ is determined by Eq.~(\ref{eq:Pij}) and will be estimated later on. Taking the average matter density of the Moon to be $\rho^{}_{\rm M} \approx 3~{\rm g}~{\rm cm}^{-3}$ and the electron fraction $Y^e_{\rm M} \approx 0.5$, we arrive at $A \approx 2.28\times 10^{-6}~{\rm eV}^2 \cdot [E/(10~{\rm MeV})]$ or equivalently $A/\Delta m^2_{21} \approx 0.03 [E/(10~{\rm MeV})]$, implying that the lunar matter effects are small even for the high-energy solar $^8{\rm B}$ neutrinos of $E \approx 10~{\rm MeV}$. In the limit of $A \ll \Delta m^2_{21} \ll \Delta m^2_{31}$, the unitary matrix $V$ given in Eq.~(\ref{eq:Vij}) can be approximately calculated, namely,
\begin{eqnarray}\label{eq:Vijapp}
V \approx \left(\begin{matrix} 1 &~ A U^*_{e1} U^{}_{e2}/\Delta m^2_{21} &~ A U^*_{e1} U^{}_{e3}/\Delta m^2_{31} \\ A U^{}_{e1} U^*_{e2}/\Delta m^2_{21} &~ 1 &~ A U^*_{e2} U^{}_{e3}/\Delta m^2_{32} \\ A U^{}_{e1} U^*_{e3}/\Delta m^2_{31}  &~ A U^{}_{e2} U^*_{e3}/\Delta m^2_{32} &~ 1 \end{matrix}\right) \; .
%     (19)
\end{eqnarray}
From Eqs.~(\ref{eq:Pij}) and (\ref{eq:Vijapp}), one can immediately derive
\begin{eqnarray}\label{eq:P12}
P^{}_{12} \approx \left(\frac{A}{\Delta m^2_{21}}\right)^2 \sin^2 2\theta^{}_{12} \sin^2 \frac{\Delta \tilde{m}^2_{21} d^{}_{\rm M}}{4E} \; .
%     (20)
\end{eqnarray}
It is now evident that the difference $P^{\rm M}_{\rm S} - P^{}_{\rm S}$ is suppressed by a factor of $A\cos 2\theta^{}_{12}/\Delta m^2_{21} \approx 0.012$, compared to the difference $P^{}_{\rm SE} - P^{}_{\rm S}$ if the matter density of the Earth is assumed to be the same as that of the Moon and the propagation distance is also equal, namely, $d^{}_{\rm M} = L^{}_{\rm E}$. The key point to understand such a difference between the lunar and terrestrial matter effects is the loss of coherence in the former case. For the same reason, the difference between $P^{\rm M}_{\rm SE}$ and $P^{}_{\rm SE}$ should also be negligible. In light of the latest neutrino oscillation data, our results demonstrate that the lunar matter effects on solar neutrinos are too small to be practically observed in realistic experiments. See Ref.~\cite{Narayan:1999ms} for an earlier discussion with different input values of $\theta^{}_{12}$ and $\Delta m^2_{21}$.

\subsection{High-Energy Astrophysical Neutrinos}

Then, to realize the coherent scenario, we consider the high-energy astrophysical neutrinos from extra-galactic sources. In fact, the lunar shadowing effects on the high-energy cosmic rays have already been detected in a number of experiments~\cite{Alexandreas:1990wj, Achard:2005az, Bartoli:2011qe, Aartsen:2013zka}, where the cosmic-ray particles can be absorbed by the Moon, reducing the flux or generating radio signals. In this subsection, we examine the matter effects induced by the Moon or the Sun on the flavor conversions of high-energy cosmic neutrinos. High-energy neutrinos are interesting since the matter term $A$ is linearly proportional to the neutrino energy, so is the oscillation length. As we will show soon, some new features of neutrino flavor conversions appear when the matter effects become remarkable.

Usually, the astrophysical neutrinos of energies above $10~{\rm TeV}$ are treated as decoherent fluxes of neutrino mass eigenstates. The reason for such a simple treatment to be valid is that the oscillation length $L^{}_{\rm osc} \approx 3.3\times 10^8~{\rm km}$, which can directly be estimated from Eq.~(\ref{eq:Losc}) for $E = 10~{\rm TeV}$ and $\Delta m^2_{21} = 7.5\times 10^{-5}~{\rm eV}^2$, is much shorter than the typical distance $D = 1~{\rm Mpc} \approx 3.1\times 10^{19}~{\rm km}$ of the astrophysical neutrino sources. This is similar to the case of solar neutrinos considered in the previous subsection. Though the IceCube observatory at the South Pole has discovered three neutrino events of energies above $1~{\rm PeV}$ by identifying the total energy deposited in the detector, it is still unclear where those neutrinos come from~\cite{Aartsen:2013bka, Aartsen:2014gkd}. Different from the case of solar neutrinos, the matter term for high-energy cosmic neutrinos induced by the Moon or the Sun can be rather large, namely, $A  \approx 0.152~{\rm eV^2}~Y^{}_e~[\rho^{}_{\rm c}/(1~{\rm g}~{\rm cm}^{-3})]~[E/{\rm TeV}]$. Therefore, the matter effects could be very significant for the high-energy cosmic neutrinos of energies above $E \gtrsim 10~{\rm TeV}$.

Although the transitional amplitude $A^{}_{ij}$ for $|\nu^{}_i\rangle \to |\nu^\prime_j\rangle$ can be exactly calculated, it is useful to derive the approximate and analytical results by taking the ratio $\Delta m_{31}^2/A \lesssim 10^{-2}\left[ {\rm TeV} /E \right]$ as a perturbation parameter, which is comparable to or even smaller than the other two parameters $\Delta m^2_{21}/\Delta m^2_{31} \approx 0.03$ and $|U^{}_{e3}|^2 \approx 0.02$. Under these approximations, the expressions in Eq.~(\ref{eq:xyz}) can be simplified to
\begin{eqnarray}\label{eq:xyzappro}
x &\approx& A \left( 1 + \frac{\Delta m^2_{31}}{A} + \frac{\Delta m^2_{31}}{A} \cdot \frac{\Delta m^2_{21}}{\Delta m^2_{31}} \right)\; , \nonumber \\
y &\approx& A^2 \left[  \left(1-|U_{e3}|^2\right) \frac{\Delta m^2_{31}}{A} + \left(1-|U_{e2}|^2\right) \frac{\Delta m^2_{31}}{A} \cdot \frac{\Delta m^2_{21}}{\Delta m^2_{31}} \right] \;, \nonumber \\
z &\approx& 1 - \frac{3}{8} \left(\frac{\Delta m^2_{31}}{A}\right)^2 \;,
%     (21)
\end{eqnarray}
and one can further obtain
\begin{eqnarray}\label{eq:xsm3yappro}
\sqrt{x^2-3y} &\approx& A \left[ 1- \frac{1}{2} \left(\frac{\Delta m^2_{31}}{A}\right) \right] \;, \nonumber \\
z+\sqrt{3(1-z^2)} &\approx& 1+\frac{3}{2} \left(\frac{\Delta m^2_{31}}{A}\right) \;, \nonumber \\
z-\sqrt{3(1-z^2)} &\approx& 1-\frac{3}{2} \left(\frac{\Delta m^2_{31}}{A}\right) \;,
%     (22)
\end{eqnarray}
where $\Delta m^2_{31}/A \approx \Delta m^2_{21}/\Delta m^2_{31} \approx |U^{}_{e3}|^2 \equiv \epsilon$ has been assumed and the higher-order terms of ${\cal O}(\epsilon^2)$ have been neglected. Note that for the neutrinos with even higher energies the ratio $\Delta m^2_{31}/A$ could be much smaller than the other two constants $\Delta m^2_{21}/\Delta m^2_{31}$ and $|U^{}_{e3}|^2$. Substituting Eqs.~(\ref{eq:xyzappro}) and (\ref{eq:xsm3yappro}) into Eq.~(\ref{eq:mtilde}), one arrives at
\begin{eqnarray}\label{eq:massappro}
\tilde{m}^2_1 \approx m_1^2 \;, \quad \tilde{m}^2_2 \approx m_3^2 \;, \quad \tilde{m}^2_3 \approx A + m_1^2  \;,
%     (23)
\end{eqnarray}
leading to the effective mass-squared differences in matter as $\Delta \tilde{m}^2_{21} \approx \Delta m^2_{31}$ and $\Delta \tilde{m}^2_{31} \approx A$ with the higher-order terms of ${\cal O}(\epsilon^2)$ neglected. In the same approximation, we can obtain the mixing matrix $V$ from Eq.~(\ref{eq:Vij}) as follows
\begin{eqnarray}\label{eq:Vappro}
V =  \left(\begin{matrix}
+U^{}_{e2}+\mathcal{O}(\epsilon) & ~\mathcal{O}(\epsilon^{1/2}) &~ U_{e1}^*+ \mathcal{O} (\epsilon ^{2}) \\
- U^{}_{e1}+\mathcal{O}(\epsilon) &~ \mathcal{O}(\epsilon^{1/2}) & ~ U_{e2}^*+\mathcal{O}(\epsilon^2) \\
\mathcal{O}(\epsilon^{3/2}) &~ 1+\mathcal{O}(\epsilon) & ~\mathcal{O}(\epsilon^{1/2}) \end{matrix}\right)
\approx\left(\begin{matrix}
+ U^{}_{e2} &~ 0 &~ U_{e1}^* \\
- U^{}_{e1} &~ 0 &~ U_{e2}^* \\
0 &~ 1 &~ 0 \end{matrix} \right)\; ,
%     (24)
\end{eqnarray}
where only the leading-order terms are retained in the last step and a proper convention for the phases has been adopted. Our numerical calculations also confirm that the order-of-magnitude estimates for the higher-order terms are correct. It is straightforward to verify the unitarity of $V$ in Eq.~(\ref{eq:Vappro}) at the level of $|U^{}_{e3}|^2$. Then, one can easily calculate the transitional amplitudes
\begin{eqnarray}\label{eq:Aijapprox}
\left[A^{}_{ij}\right]
&=& \left(\begin{matrix} |U^{}_{e1}|^2 e^{-{\rm i}\tilde{\varphi}^{}_{31}} + |U^{}_{e2}|^2 + \mathcal{O}(\epsilon) &~ U^{}_{e1} U^*_{e2} \left(e^{-{\rm i}\tilde{\varphi}^{}_{31}} - 1\right)+\mathcal{O}(\epsilon) &~ \mathcal{O}(\epsilon^{1/2}) \cr U^*_{e1} U^{}_{e2} \left(e^{-{\rm i}\tilde{\varphi}^{}_{31}} - 1\right) + \mathcal{O}(\epsilon) &~ |U^{}_{e2}|^2 e^{-{\rm i}\tilde{\varphi}^{}_{31}} + |U^{}_{e1}|^2+\mathcal{O}(\epsilon) &~ \mathcal{O}(\epsilon^{1/2}) \cr \mathcal{O}(\epsilon^{1/2}) & \mathcal{O}(\epsilon^{1/2}) &~ e^{-{\rm i}\tilde{\varphi}^{}_{21}}+\mathcal{O}(\epsilon) \end{matrix}\right) \; \nonumber\\
&\approx & \left(\begin{matrix} |U^{}_{e1}|^2 e^{-{\rm i}\tilde{\varphi}^{}_{31}} + |U^{}_{e2}|^2 &~ U^{}_{e1} U^*_{e2} \left(e^{-{\rm i}\tilde{\varphi}^{}_{31}} - 1\right) &~ 0 \cr U^*_{e1} U^{}_{e2} \left(e^{-{\rm i}\tilde{\varphi}^{}_{31}} - 1\right) &~ |U^{}_{e2}|^2 e^{-{\rm i}\tilde{\varphi}^{}_{31}} + |U^{}_{e1}|^2 &~ 0 \cr 0 &~ 0 &~ e^{-{\rm i}\tilde{\varphi}^{}_{21}}\end{matrix}\right) \; ,
%     (25)
\end{eqnarray}
where the oscillation phases $\tilde{\varphi}^{}_{ij} \equiv \Delta \tilde{m}^2_{ij} d^{}_{\rm M}/(2E)$ for $ij = 31, 21$ have been defined. From Eq.~(\ref{eq:Aijapprox}), one can further find out all the nonzero transitional probabilities
\begin{eqnarray}\label{eq:Pijapprox}
P^{}_{11} \approx P^{}_{22} \approx 1 - 4|U^{}_{e1}|^2 |U^{}_{e2}|^2 \sin^2\left(\tilde{\varphi}^{}_{31}/2\right) \; , \quad
P^{}_{12} \approx P^{}_{21} \approx 4|U^{}_{e1}|^2 |U^{}_{e2}|^2 \sin^2\left(\tilde{\varphi}^{}_{31}/2\right) \; ,
%     (26)
\end{eqnarray}
and $P^{}_{33} \approx 1$. As implied by Eq.~(\ref{eq:massappro}), $\tilde{\varphi}^{}_{31}/2 = \Delta \tilde{m}^2_{31} d^{}_{\rm M}/(4E) \approx A d^{}_{\rm M}/(4E) = G^{}_{\rm F} N^{}_e d^{}_{\rm M}/\sqrt{2}$ is independent of the neutrino energy $E$, and proportional to the traveling distance $d^{}_{\rm M}$ and the net electron number density $N^{}_e$. Consequently, the typical size $d^{}_{\rm M}$ of the astrophysical object is crucially important for the transitional probability $P^{}_{12}$ to be significant.

In order to compute the probabilities $\widehat{P}^{\rm coh}_{\alpha \beta}$ in the coherent scenario, we have to estimate the interference terms $\widehat{I}^{}_{\alpha \beta}$ in Eq.~(\ref{eq:Iab}), which depend on both the transitional amplitudes $A^{}_{ij}$ in Eq.~(\ref{eq:Aijapprox}) and the distance $L$ between the intermediate astrophysical object and the detector. Taking the Moon or the Sun as the intermediate astrophysical object, we analyze the corresponding matter effects on the flavor conversions of high-energy cosmic neutrinos.

\subsubsection{The Lunar Case}

The average distance between the Moon and the Earth is $L^{}_{\rm ME} = 3.84\times 10^5~{\rm km}$, which is much shorter than the neutrino oscillation lengths in vacuum $L^{21}_{\rm osc} \approx 3.3\times 10^8~{\rm km}~[E/(10~{\rm TeV})]$ and $L^{31}_{\rm osc} \approx 9.9\times 10^6~{\rm km}~[E/(10~{\rm TeV})]$, corresponding respectively to two neutrino mass-squared differences $\Delta m^2_{21} = 7.5\times 10^{-5}~{\rm eV}^2$ and $\Delta m^2_{31} = 2.5\times 10^{-3}~{\rm eV}^2$. As a consequence, the oscillation phase developed during the propagation between the Moon and the Earth is negligible. On the other hand, the diameter of the Moon is $d^{}_{\rm M} \approx 3.48\times 10^3~{\rm km}$, so the oscillation phase inside the Moon can be estimated to be $\tilde{\varphi}^{}_{31}/2 \approx 0.67~Y^{}_e~[\rho^{}_{\rm M}/(1~{\rm g}~{\rm cm}^{-3})]$, with which a considerable transitional probability $P^{}_{21}$ could be derived from Eq.~(\ref{eq:Pijapprox}). However, the final probabilities $\widehat{P}^{\rm coh}_{\alpha \beta}$ will also depend on the interference terms
\begin{eqnarray}\label{eq:IabMoon}
\widehat{I}^{}_{\alpha \beta} &=& \sum^3_{i = 1} k^\alpha_i \sum_{j > k} 2 {\rm Re}\left\{ U^{}_{\beta j} U^*_{\beta k} A^{}_{ij} A^*_{ik} \exp\left[-{\rm i}\Delta m^2_{jk} L^{}_{\rm ME}/(2E)\right]\right\} \nonumber \\
&\approx& \sum^3_{i = 1} k^\alpha_i \sum_{j > k} 2 {\rm Re}\left( U^{}_{\beta j} U^*_{\beta k} A^{}_{ij} A^*_{ik} \right) \; ,
%     (6)
\end{eqnarray}
where it should be noticed that $L^{}_{\rm ME}$ is much shorter than the oscillation lengths $L^{21}_{\rm osc}$ and $L^{31}_{\rm osc}$ in vacuum. Using Eq~(\ref{eq:Pab}), after some straightforward calculations, one can further show that the decoherent probability $\widehat{P}^{\rm dec}_{\alpha \beta}$ and the interference term $\widehat{I}^{}_{\alpha \beta}$ will cancel each other, leading to
\begin{eqnarray}\label{eq:PcohMoon}
\widehat{P}^{\rm coh}_{\alpha \beta} \approx P_{\alpha \beta} =\sum^3_{i=1} k^{\alpha}_i |U^{}_{\beta i}|^2 \;
%     (26)
\end{eqnarray}
at the zeroth order. Hence the matter effects induced by the Moon will always be insignificant for high-energy cosmic neutrinos.

The above conclusions can be reached in a more transparent way. Given the fact that the distance between the Moon and the Earth is considerably smaller than the oscillation lengths, the situation should be equivalent to that of placing the detector just on the surface of the Moon. Furthermore, with the help of Eq.~(\ref{eq:Vappro}), one can find the effective mixing matrix $\tilde{U} = U V$ which relates the mass eigenstates in matter to the flavor eigenstates:
\begin{eqnarray}\label{eq:Umatrix}
\tilde{U} = U V &\approx& \left(\begin{matrix}
0 &~ 0 &~ 1 \\
U_{\mu 1}U_{e2} - U_{\mu 2}U_{e1}&~ U_{\mu 3} &~ 0 \\
U_{\tau 1}U_{e2} - U_{\tau 2}U_{e1}&~                                                                                     U_{\tau 3} &~ 0 \end{matrix}\right) \;,
%     (29)
\end{eqnarray}
where the higher-order terms of $\mathcal{O}(\epsilon^{1/2})$ have been omitted as in the last step in Eq.~(\ref{eq:Vappro}). The effective mass eigenvalues have been given in Eq.~(\ref{eq:massappro}). In the antineutrino case, one can just perform the replacements $A \to - A$ and $U \to U^*$ in the effective mass eigenvalues and in the effective mixing matrix. The form of matrix $\tilde{U}$ can be actually obtained in a simpler and more accurate way. Since the effective Hamiltonian can be diagonalized as in Eq.~(\ref{eq:diag}), we have
\begin{eqnarray}\label{eq:diagapprox1}
V^\dagger H^{}_{\rm m} V &=& \frac{1}{2E} \left[ \tilde{U}^\dagger U \left(\begin{matrix} m^2_1 & 0 & 0 \\ 0 & m^2_2 & 0 \\ 0 & 0 & m^2_3 \end{matrix}\right) U^\dagger \tilde{U} + \tilde{U}^\dagger  \left(\begin{matrix} A & 0 & 0 \\ 0 & 0 & 0 \\ 0 & 0 & 0 \end{matrix}\right) \tilde{U} \right] \; \nonumber \\
&=& \frac{A}{2E} \left[ \frac{\Delta m^2_{31}}{A}~\tilde{U}^\dagger U \left(\begin{matrix} 0 & 0 & 0 \\ 0 & \displaystyle \frac{\Delta m^2_{21}}{\Delta m^2_{31}} & 0 \\ 0 & 0 & 1 \end{matrix}\right) U^\dagger \tilde{U} + \tilde{U}^\dagger  \left(\begin{matrix} 1 & 0 & 0 \\ 0 & 0 & 0 \\ 0 & 0 & 0 \end{matrix}\right) \tilde{U} \right] \;,
\end{eqnarray}
which is a diagonal matrix. Therefore, the last term in the second line of Eq.~(\ref{eq:diagapprox1}), namely,
\begin{eqnarray}\label{eq:diagapprox2}
 \tilde{U}^\dagger  \left(\begin{matrix} 1 &~ 0 &~ 0 \\ 0 &~ 0 &~ 0 \\ 0 &~ 0 &~ 0 \end{matrix}\right) \tilde{U} = \left(\begin{matrix} |\tilde{U}_{e1}|^2 &~ \tilde{U}^{*}_{e1}\tilde{U}^{}_{e2} &~ \tilde{U}^{*}_{e1}\tilde{U}^{}_{e3} \\ \tilde{U}^{}_{e1}\tilde{U}^{*}_{e2} &~ |\tilde{U}^{}_{e2}|^2 &~ \tilde{U}^{*}_{e2}\tilde{U}_{e3} \\ \tilde{U}_{e1}\tilde{U}^{*}_{e3} &~ \tilde{U}^{}_{e2}\tilde{U}^{*}_{e3} &~ |\tilde{U}_{e3}|^2 \end{matrix}\right) \;,
\end{eqnarray}
should be diagonal up to ${\cal O}(\epsilon)$. This can be achieved if and only if two of the matrix elements $\tilde{U}^{}_{ei}$ (for $i=1, 2, 3$) are vanishing. This is the case for Eq.~(\ref{eq:Umatrix}). One can explicitly check that $|\nu^{}_{e}\rangle$ can always be identified as the mass eigenstate with largest eigenvalue of $A+m^2_1$, while the other two states $|\nu^{}_{\mu}\rangle$ and $|\nu^{}_{\tau}\rangle$ can oscillate from one to another with the relevant neutrino mass-squared difference $\Delta m^2_{31}$. It is worthwhile to note that these arguments are justified as long as $\Delta m_{31}^2/A \approx 10^{-3}\left[ {\rm 10~TeV} /E \right] \ll 1$ is satisfied.

Now it is evident that the electron neutrino state coincides with the heaviest mass eigenstate in matter, and only the flavor conversions between $|\nu^{}_{\mu}\rangle$ and $|\nu^{}_{\tau}\rangle$ take place. For the latter, the relevant neutrino mass-squared difference is given by $\Delta \tilde{m}^2_{21} \approx \Delta m^2_{31}$, so the corresponding oscillation length $L^{31}_{\rm osc} \approx 9.9\times 10^6~{\rm km}~[E/(10~{\rm TeV})]$ is much longer than the diameter of the Moon. Consequently, the oscillation phase for $|\nu^{}_\mu\rangle \to |\nu^{}_\tau\rangle$ will never develop significantly while the overall phase for $| \nu^{}_e \rangle$ is undetectable. We have confirmed numerically that the matter effects induced by the Moon on the flavor conversions of high-energy cosmic neutrinos are as small as $0.1\%$, and thus can be ignored. This conclusion is also applicable to the case when the high-energy neutrinos traverse the Earth before arriving in the detector.

\subsubsection{The Solar Case}
From the previous discussions, we have seen that the distance between the astrophysical object and the Earth is very important. As for the Sun, the average distance to the Earth is $L^{}_{\rm SE} = 1.5\times 10^8~{\rm km}$, which is comparable to $L^{21}_{\rm osc}$ and much larger than $L^{31}_{\rm osc}$ when the neutrino energy is around $10~{\rm TeV}$. However, for even higher neutrino energies $E \gtrsim 10^3~{\rm TeV}$,  we find $L^{21}_{\rm osc} \gtrsim 3.3\times 10^{10}~{\rm km}$ and $L^{31}_{\rm osc} \gtrsim 9.9\times 10^8~{\rm km}$, indicating that the matter effects induced by the Sun will be negligible just like in the lunar case. For this reason, we concentrate on the neutrino energies below $10^3~{\rm TeV}$.

Given the Sun's diameter $d_{\rm S} \approx 1.4 \times 10^6~{\rm km}$, one can obtain the relevant oscillation phase $\tilde{\varphi}^{}_{31}/2 \approx 270~Y^{}_e~[\rho^{}_{\rm c}/(1~{\rm g}~{\rm cm}^{-3})]$, which is very large for $Y^{}_e \approx 0.67$ and $\rho^{}_{\rm c} = 150~{\rm g}~{\rm cm}^{-3}$ in the solar core. Therefore, the nonzero transitional probabilities in Eq.~(\ref{eq:Pijapprox}) should be averaged over many cycles of oscillations due to the variations of the traveled distance in matter, namely,
\begin{eqnarray}\label{eq:Pijaverage}
P^{}_{11} \approx P^{}_{22} \approx 1 - 2|U^{}_{e1}|^2 |U^{}_{e2}|^2 \; , \quad
P^{}_{12} \approx P^{}_{21} \approx 2|U^{}_{e1}|^2 |U^{}_{e2}|^2 \; ,\quad
P^{}_{33} \approx 1 \;.
%     (26)
\end{eqnarray}
 The interference terms read
\begin{eqnarray}\label{eq:IabSun}
\widehat{I}^{}_{\alpha \beta} &=& \sum^3_{i = 1} k^\alpha_i \sum_{j > k} 2 {\rm Re}\left\{ U^{}_{\beta j} U^*_{\beta k} A^{}_{ij} A^*_{ik} \exp\left[-{\rm i}\Delta m^2_{jk} L^{}_{\rm SE}/(2E)\right]\right\} \; \nonumber \\
&\approx&  \sum^3_{i = 1} k^\alpha_i 2 {\rm Re}\left\{ U^{}_{\beta 2} U^*_{\beta 1} A^{}_{i2} A^*_{i1} \exp\left[-{\rm i}\Delta m^2_{21} L^{}_{\rm SE}/(2E)\right]\right\}  \; ,
%     (6)
\end{eqnarray}
where the approximation is validated according to  Eq.~(\ref{eq:Aijapprox}). The oscillation phase $\tilde{\varphi}^{}_{31}/2$ in the transition amplitudes $A^{}_{ij}$ is large and will be eventually averaged out when one integrate the observation angle, so the final results would not depend on this matter-related phase. 

In the actual calculations, we assume that the matter density of the Sun is constant and take its average value $\rho^{}_{\rm S} \approx 1.408~{\rm g}~{\rm cm}^{-3}$ with $Y^{}_e \approx 0.7$. Such an approximation is good enough for the high-energy neutrinos of $E \gtrsim 10~{\rm TeV}$ for the following reasons:
\begin{itemize}
\item As long as the condition $\Delta m_{31}^2/A \ll 1$ is satisfied, the transition probabilities will be given by those in Eq.~(\ref{eq:Pijaverage}) and the interference terms should also be averaged. In addition, no matter what exactly the solar density profile is, the mixing matrix $V$ is fixed as in Eq.~(\ref{eq:Vappro}), so varying the matter density profile can only have a minor impact on the oscillation phase $\tilde{\varphi}^{}_{31}/2$. When neutrinos are propagating inside the Sun, the oscillation phase $\tilde{\varphi}^{}_{31}/2$ will be integrated along the neutrino trajectory. The final value of this phase is determined by the trajectory-averaged matter density, namely, $\tilde{\varphi}^{}_{31}/2 \approx 6750~Y^{}_e~[\bar{\rho}^{}_{\rm c}/(25~{\rm g}~{\rm cm}^{-3})]~[d/ d^{}_{\rm S}]$ with $\bar{\rho}^{}_{\rm c}$ being the averaged density along the trajectory of the length of $d$. In practice, the exact value of $\tilde{\varphi}^{}_{31}/2$ will be unimportant, since it depends strongly on the observation angle from the detector at the Earth. For instance, when averaged over the solid angle covered by the Sun, $\tilde{\varphi}^{}_{31}/2$ can vary from $0$ to $6750$, indicating that the oscillation phase will be eventually averaged out and the matter density profile is not quite relevant. The assumption of a constant matter density $\rho^{}_{\rm S} \approx 1.408~{\rm g}~{\rm cm}^{-3}$ with $Y^{}_e \approx 0.7$ is thus justified for the Sun.
\item Then, it is important to know whether the condition $\Delta m_{31}^2/A \ll 1$ can always be satisfied inside the Sun. Only in the region of $R \gtrsim 0.995~R^{}_{\rm S}$ with $R_{\rm S}$ being the solar radius, one can actually find $\Delta m_{31}^2/A \gtrsim 0.03$. The traveling distance of neutrinos in this region is about $7\times 10^3~{\rm km}$, whereas the corresponding oscillation length should be at least $3 \times 10^{5}~{\rm km}$. Therefore, the neutrino state remains almost unchanged when propagating through this particular region. 
\end{itemize}

To illustrate the solar matter effects on high-energy cosmic neutrinos, we consider two distinct neutrino production mechanisms in the astrophysical sources (e.g., GRBs and AGN), where the accelerated protons will interact with ambient protons ($pp$) or photons ($p\gamma$), copiously producing $\pi^+$'s (or $\pi^-$'s) that further decay into $\mu^+$'s (or $\mu^-$'s) and $\nu^{}_\mu$'s (or $\overline{\nu}^{}_\mu$'s). If the magnetic field in the sources is strong, $\mu^+$ and $\mu^-$ will loose rapidly their energies via synchrotron radiation before decaying into secondary neutrinos and antineutrinos~\cite{Winter:2012xq}. For this kind of $\mu$-damped sources, one may not expect the presence of electron (anti)neutrinos. Otherwise, we have $\mu^+ \to e^+ + \nu^{}_e + \overline{\nu}^{}_\mu$ and $\mu^- \to e^- + \overline{\nu}^{}_e + \nu^{}_\mu$. Therefore, the flavor compositions for high-energy cosmic neutrinos at the sources and those at the detectors can be summarized as below:
\begin{itemize}
\item \emph{pp sources}
\begin{eqnarray}\label{eq:pp}
\left\{\phi^{\rm S}_{\nu_{e}}, \phi^{\rm S}_{\nu_{\mu}}, \phi^{\rm S}_{\nu_{\tau}}, \phi^{\rm S}_{\overline{\nu}_{e}}, \phi^{\rm S}_{\overline{\nu}_{\mu}}, \phi^{\rm S}_{\overline{\nu}_{\tau}}\right\} &=& \phi \left\{\frac{1}{6}, \frac{1}{3}, 0, \frac{1}{6}, \frac{1}{3}, 0\right\} \;, \nonumber \\
\left\{\phi^{\rm S}_{e}, \phi^{\rm S}_{\mu}, \phi^{\rm S}_{\tau} \right\} &=& \phi \left\{\frac{1}{3}, \frac{2}{3}, 0\right\} \;, \nonumber \\
\left\{\phi^{\rm D}_{e}, \phi^{\rm D}_{\mu}, \phi^{\rm D}_{\tau} \right\} &=& \phi \left\{0.313, 0.346, 0.341\right\} \;, %
\end{eqnarray}

\item \emph{$\mu$-damped pp sources}
\begin{eqnarray}\label{eq:ppd}
\left\{\phi^{\rm S}_{\nu_{e}}, \phi^{\rm S}_{\nu_{\mu}}, \phi^{\rm S}_{\nu_{\tau}}, \phi^{\rm S}_{\overline{\nu}_{e}}, \phi^{\rm S}_{\overline{\nu}_{\mu}}, \phi^{\rm S}_{\overline{\nu}_{\tau}} \right\} &=& \phi \left\{0, \frac{1}{2}, 0, 0, \frac{1}{2}, 0\right\} \;, \nonumber \\
\left\{\phi^{\rm S}_{e}, \phi^{\rm S}_{\mu}, \phi^{\rm S}_{\tau} \right\} &=& \phi \left\{0, 1, 0\right\}\;, \nonumber \\
\left\{\phi^{\rm D}_{e}, \phi^{\rm D}_{\mu}, \phi^{\rm D}_{\tau} \right\} &=& \phi \left\{0.195, 0.422, 0.383\right\} \;; %
\end{eqnarray}

\item \emph{p$\gamma$ sources}
\begin{eqnarray}\label{eq:pg}
\left\{\phi^{\rm S}_{\nu_{e}}, \phi^{\rm S}_{\nu_{\mu}}, \phi^{\rm S}_{\nu_{\tau}}, \phi^{\rm S}_{\overline{\nu}_{e}}, \phi^{\rm S}_{\overline{\nu}_{\mu}}, \phi^{\rm S}_{\overline{\nu}_{\tau}} \right\} &=& \phi \left\{\frac{1}{3}, \frac{1}{3}, 0, 0, \frac{1}{3}, 0\right\}\;, \nonumber \\
\left\{\phi^{\rm S}_{e}, \phi^{\rm S}_{\mu}, \phi^{\rm S}_{\tau} \right\} &=& \phi \left\{\frac{1}{3}, \frac{2}{3}, 0\right\} \;, \nonumber \\
\left\{\phi^{\rm D}_{e}, \phi^{\rm D}_{\mu}, \phi^{\rm D}_{\tau} \right\} &=& \phi \left\{0.313, 0.346, 0.341\right\} \;; %
\end{eqnarray}

\item \emph{$\mu$-damped p$\gamma$ sources}
\begin{eqnarray}\label{eq:pgd}
\left\{\phi^{\rm S}_{\nu_{e}}, \phi^{\rm S}_{\nu_{\mu}}, \phi^{\rm S}_{\nu_{\tau}}, \phi^{\rm S}_{\overline{\nu}_{e}}, \phi^{\rm S}_{\overline{\nu}_{\mu}}, \phi^{\rm S}_{\overline{\nu}_{\tau}}\right\} &=& \phi \left\{0, 1, 0, 0, 0, 0\right\} \;, \nonumber \\
\left\{\phi^{\rm S}_{e}, \phi^{\rm S}_{\mu}, \phi^{\rm S}_{\tau} \right\} &=& \phi \left\{0, 1, 0\right\} \;, \nonumber \\
\left\{\phi^{\rm D}_{e}, \phi^{\rm D}_{\mu}, \phi^{\rm D}_{\tau} \right\} &=& \phi \left\{0.195, 0.422, 0.383 \right\} \; . %
\end{eqnarray}

\end{itemize}

Some comments on the neutrino flavor compositions in the above four scenarios are useful. First, $\phi^{\rm S}_{\nu^{}_{\alpha}}$ and $\phi^{\rm S}_{\overline{\nu}^{}_{\alpha}}$ denote the original fluxes of  $\nu^{}_{\alpha}$ and $\overline{\nu}^{}_{\alpha}$ for $\alpha = e, \mu, \tau$ at the sources, respectively. In addition, $\phi^{\rm S}_{\alpha} \equiv \phi^{\rm S}_{\nu^{}_\alpha} + \phi^{\rm S}_{\overline{\nu}^{}_{\alpha}}$ is the sum of the $\nu^{}_\alpha$ and $\overline{\nu}^{}_\alpha$ fluxes, while $\phi = \phi^{\rm S}_e + \phi^{\rm S}_\mu + \phi^{\rm S}_\tau$ stands for the total flux. Note that we distinguish between the original fluxes of $\nu^{}_\alpha$ and $\overline{\nu}^{}_\alpha$, since they may have different oscillation behaviors when the matter effects become relevant. At the detector, after neutrino oscillations, the final fluxes are given by~\cite{Learned:1994wg, Farzan:2002ct, Serpico:2005bs, Xing:2006uk, Winter:2006ce, Xing:2006xd}
\begin{eqnarray}\label{eq:phidstd}
\phi^{\rm D}_\alpha = \sum^\tau_{\beta = e} \phi^{\rm S}_\beta P^{}_{\beta \alpha}  = \sum^3_{i=1} \sum^\tau_{\beta = e} \phi^{\rm S}_\beta |U^{}_{\beta i}|^2 |U^{}_{\alpha i}|^2  \; ,
\end{eqnarray}
where Eq.~(\ref{eq:PS}) with $k^\beta_i = |U^{}_{\beta i}|^2$ has been used. As the neutrino telescopes, such as IceCube~\cite{Ahrens:2003ix} and KM3NeT~\cite{Adrian-Martinez:2016fdl}, cannot distinguish neutrinos and antineutrinos~\footnote{It is certainly possible to discriminate between $\overline{\nu}^{}_e$ and the others via the Glashow resonance $\overline{\nu}^{}_e + e^- \to W^- \to$ anything in neutrino telescopes, when the $\overline{\nu}^{}_e$ energy exceeds the threshold $E^{}_{\overline{\nu}^{}_e} \approx 6.3~{\rm PeV}$~\cite{Glashow:1960zz, Berezinsky:1977sf, Xing:2011zm, Bhattacharya:2011qu}. However, we tentatively ignore this possibility in this work.}, we simply sum over the fluxes of neutrinos and antineutrinos at the detector $\phi^{\rm D}_{\alpha} = \phi^{\rm D}_{\nu^{}_\alpha} + \phi^{\rm D}_{\overline{\nu}^{}_\alpha}$ in the absence of matter effects. Second, we assume the normal neutrino mass ordering and take the best-fit values of three mixing angles $\theta^{}_{12} = 33.62^{\circ}$, $\theta^{}_{13} = 8.54^{\circ}$, and $\theta^{}_{23} = 47.2^{\circ}$ and the CP-violating phase $\delta=234^{\circ}$ from the latest global-fit analysis of neutrino oscillation data. Third, regarding the $pp$ and $p\gamma$ collisions in the astrophysical sources, we assume that the equal amounts of $\pi^+$'s and $\pi^-$'s are produced in the former case and only $\pi^+$'s are generated in the latter case. See, e.g., Ref.~\cite{Xing:2011zm}, for a discussion about the impact on the neutrino flavor composition if this assumption is relaxed.

In order to incorporate the solar matter effects in a realistic detection, we have to consider the exact distance $d(\Theta)$ traveled by neutrinos and antineutrinos inside the Sun and the distance $L(\Theta)$ in vacuum between the Sun and the Earth, where $\Theta$ is the zenith angle between the trajectory and the line connecting the solar center and the detector. Therefore, the averaged flux of neutrinos and antineutrinos at the detector within the solid angle ${\rm d}\Omega \equiv 2\pi {\rm d}(\cos \Theta)$ can be calculated as
\begin{eqnarray}\label{eq:avePhi}
\widehat{\Phi}^{\rm D}_\alpha \equiv \int {\rm d}\Omega \frac{{\rm d}}{{\rm d} \Omega} \left[\sum^\tau_{\beta = e} \widehat{P}^{\rm coh}_{\beta \alpha}(\Theta) \phi^{\rm S}_{\nu^{}_\beta}\right] + \int {\rm d}\Omega \frac{{\rm d}}{{\rm d} \Omega} \left[\sum^\tau_{\beta = e} \widehat{P}^{{\rm coh}}_{\overline{\beta} \overline{\alpha}} (\Theta) \phi^{\rm S}_{\overline{\nu}^{}_\beta}\right] \; ,
\end{eqnarray}
where $\widehat{P}^{\rm coh}_{\beta \alpha}(\Theta)$ and $\widehat{P}^{\rm coh}_{\overline{\beta} \overline{\alpha}}(\Theta)$ stand for the transitional probabilities for $\nu^{}_\beta \to \nu^{}_\alpha$ and $\overline{\nu}^{}_\beta \to \overline{\nu}^{}_\alpha$, respectively. It is straightforward to figure out $d(\Theta) = d^{}_{\rm S} \cos \left[\arcsin(2L^{}_{\rm SE} \sin\Theta/d^{}_{\rm S})\right]$ and $L(\Theta) = L^{}_{\rm SE} \cos \Theta - d(\Theta)/2$, where $L^{}_{\rm SE}$ is the distance from the solar center to the detector at the Earth. As $L^{}_{\rm SE}$ is much longer than the solar radius, the maximal zenith angle is $\Theta^{}_{\rm max} \approx d^{}_{\rm S}/(2L^{}_{\rm SE}) \approx 0.27^\circ$, so the distance $d(\Theta)$ traveled by neutrinos and antineutrinos inside the Sun varies from $d^{}_{\rm S}$ to 0, which justifies the averaged probabilities over the variations of the distance in Eq.~(\ref{eq:Pijaverage}). On the other hand, the tiny angle $\Theta^{}_{\rm max}$ is much smaller than the current angular resolution of neutrino events in neutrino telescopes, implying that the introduction of the averaged flux in Eq.~(\ref{eq:avePhi}) is relevant and necessary.  The solid angle spanned by the Sun relative to the detector is $\Omega \approx 7\times 10^{-5}$, so it should be very challenging for the present neutrino telescope, such as IceCube, to register enough neutrino events in order to probe the solar matter effects.
%%%%%%%%%%%%%%%%%%%%%%%%%%%%%%%% Fig. 2 %%%%%%%%%%%%%%%%%%%%%%%%%%%%%%%%%
\begin{figure}[t!]
\begin{center}
\subfigure{
\hspace{-0.2cm}
\includegraphics[width=0.45\textwidth]{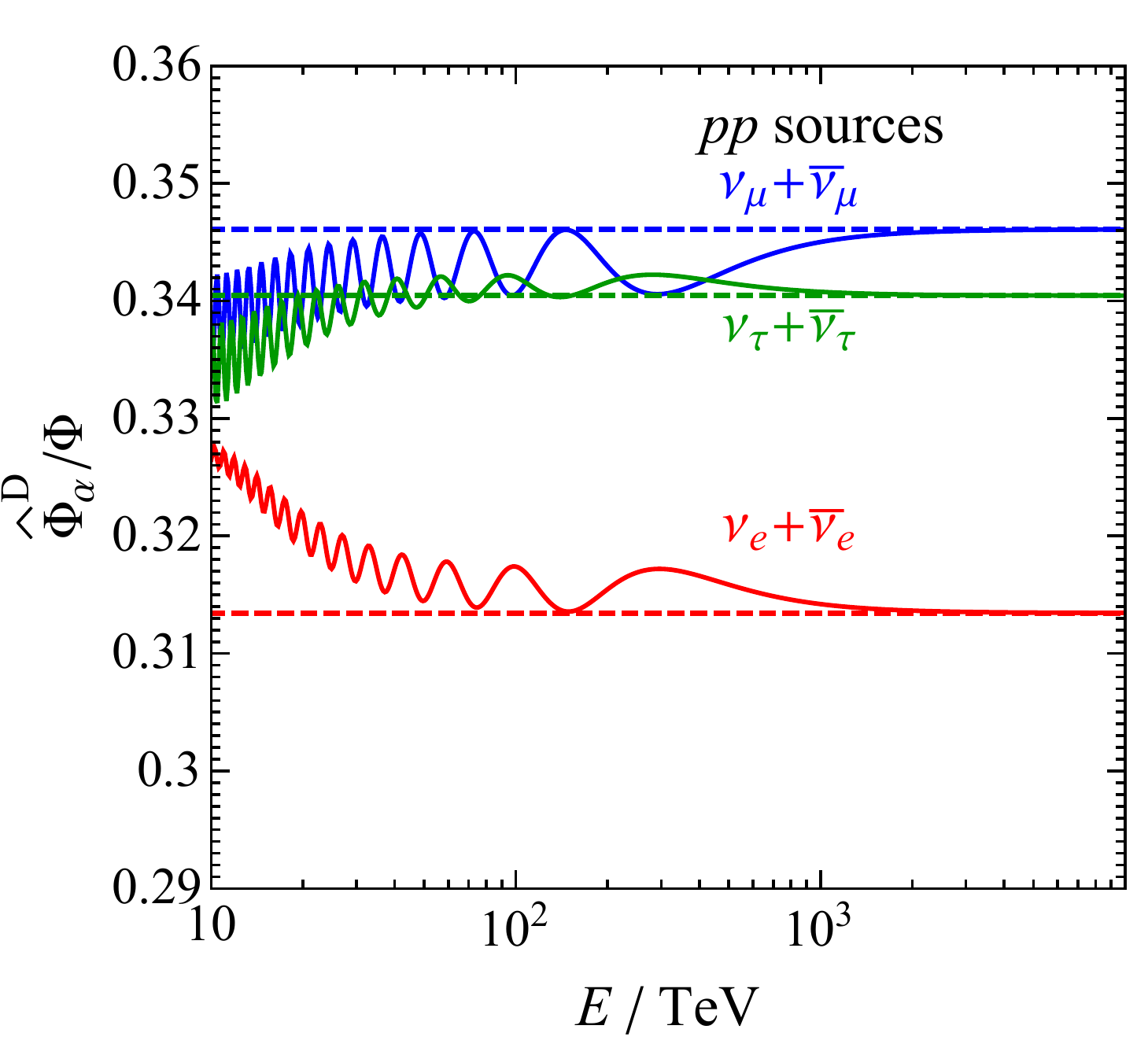} }
\subfigure{
\includegraphics[width=0.45\textwidth]{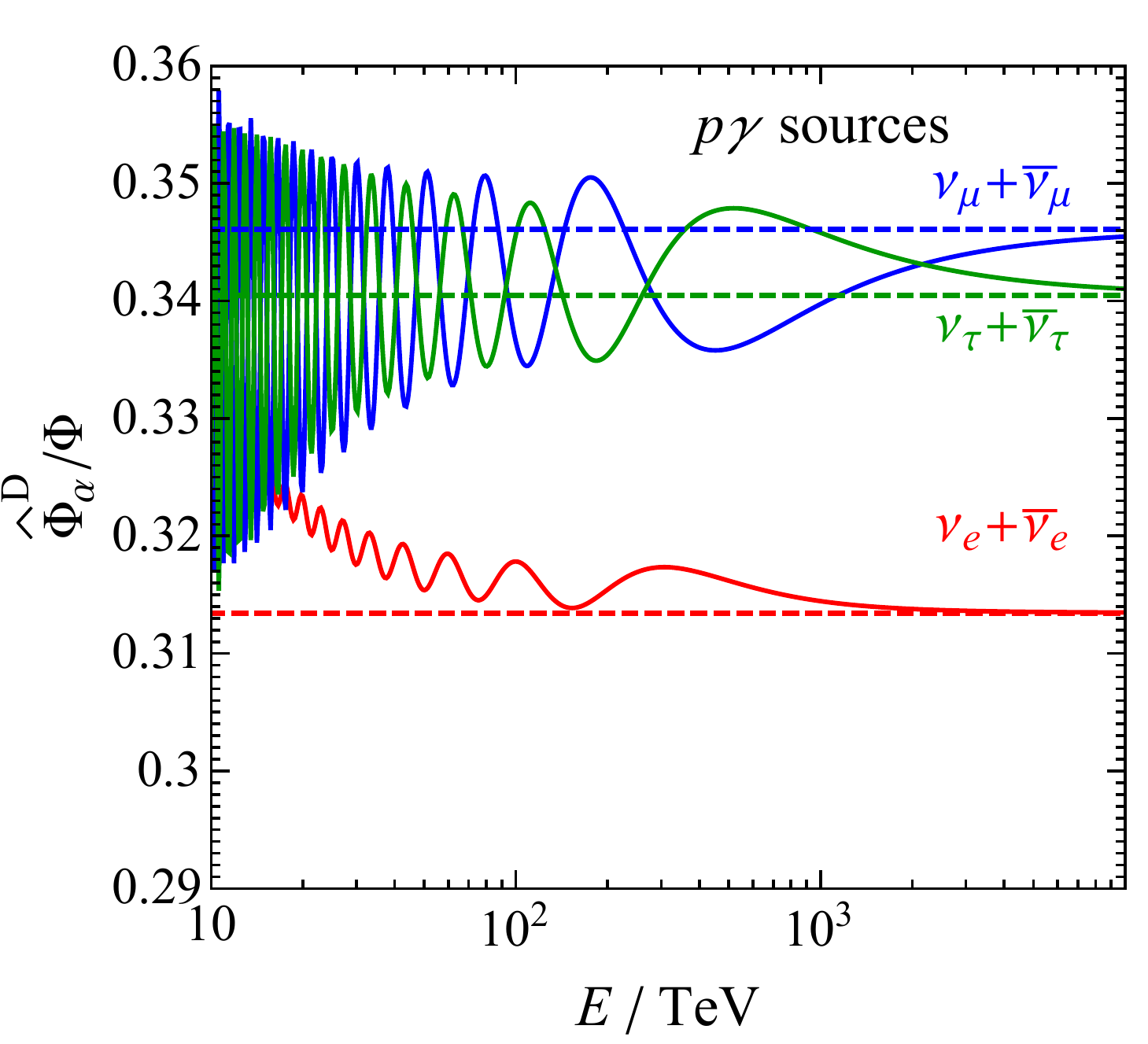} }
\subfigure{
\hspace{-0.2cm}
\includegraphics[width=0.45\textwidth]{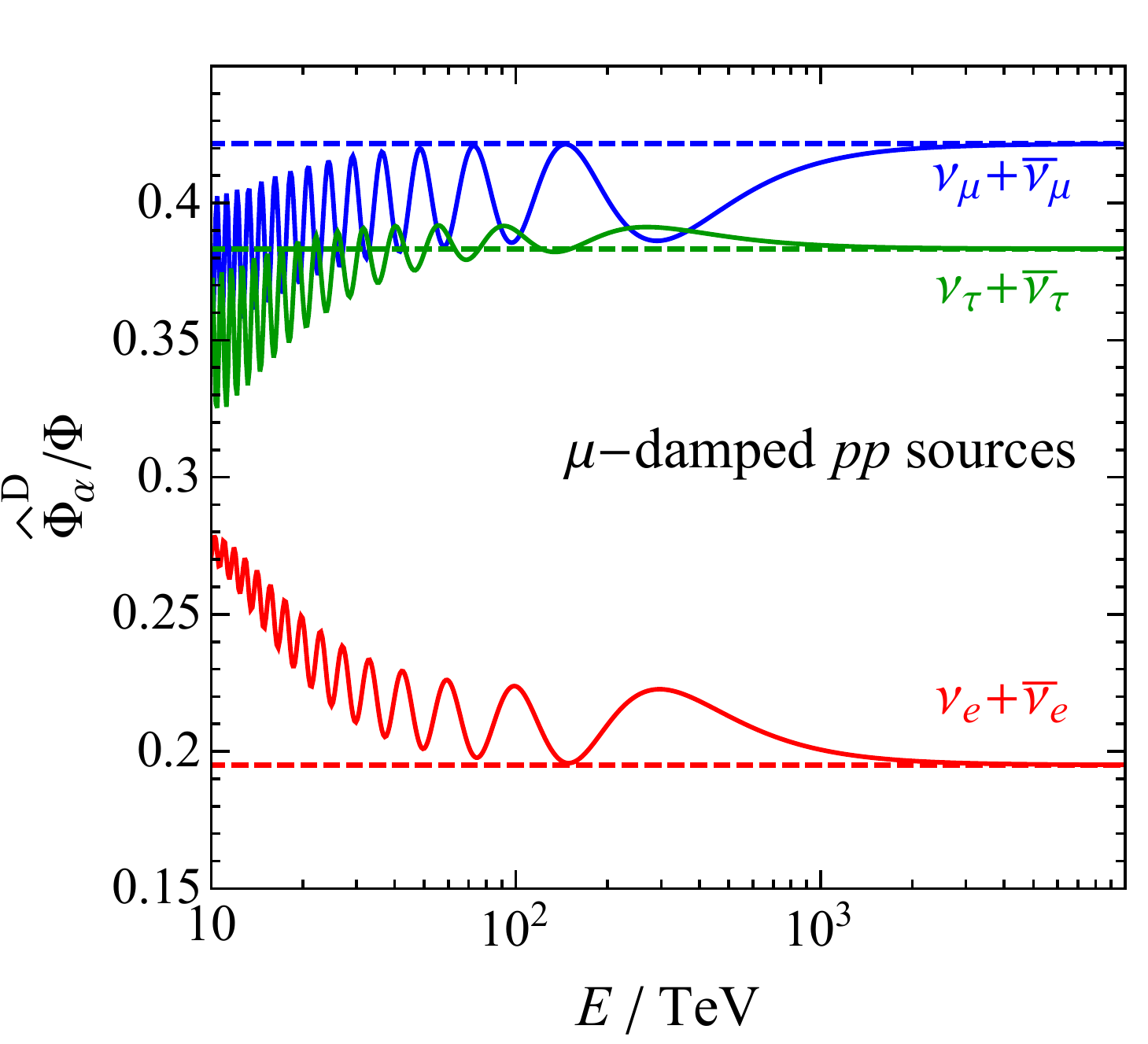} }
\subfigure{
\includegraphics[width=0.45\textwidth]{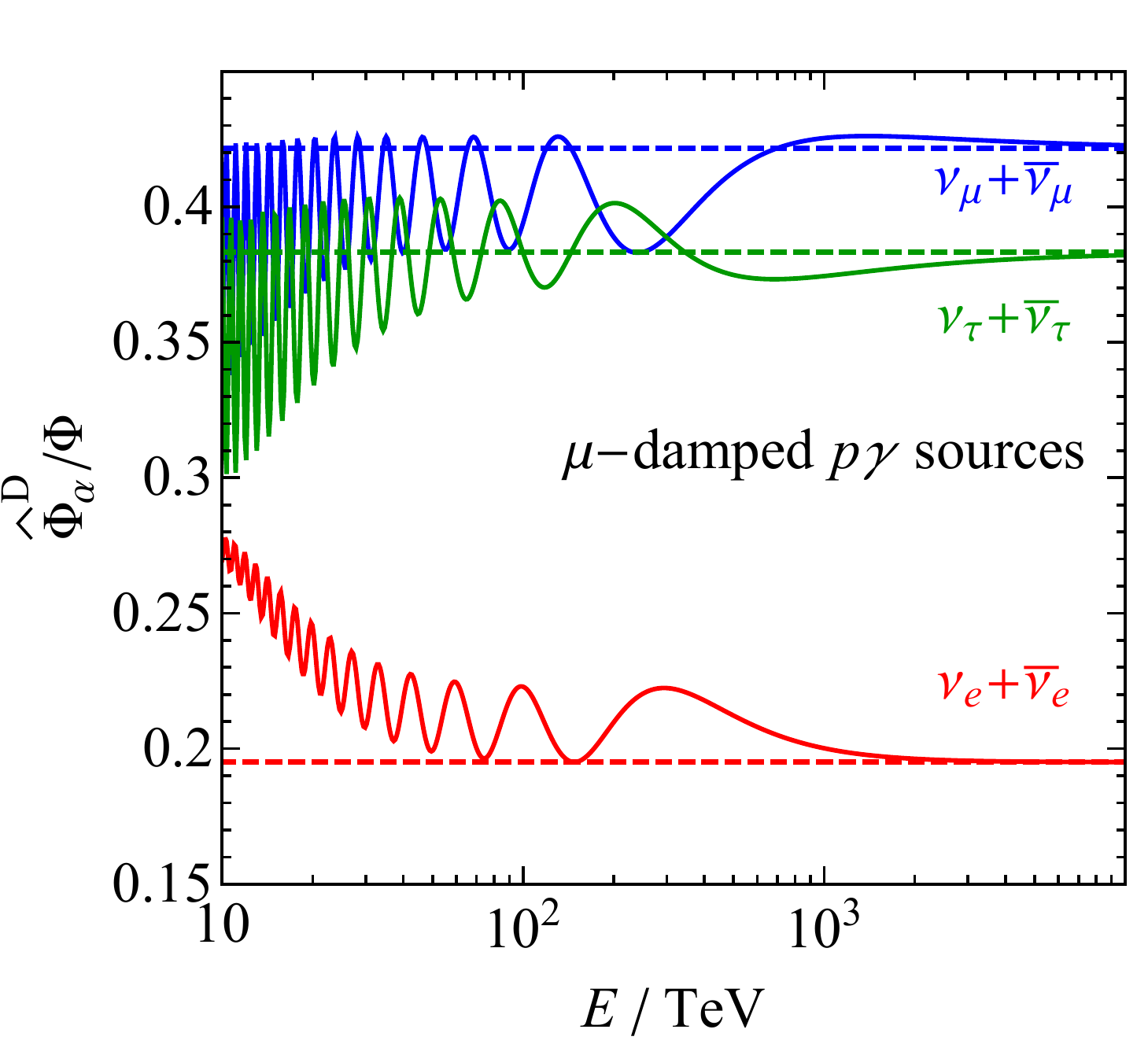} }
\end{center}
\vspace{-0.5cm}
\caption{The flavor ratios $\widehat{\Phi}^{\rm D}_\alpha/\Phi$ of high-energy cosmic neutrinos after the solar matter effects are taken into account, where the dashed lines stand for the results when the solar matter effects are omitted. The best-fit values of neutrino mixing angles and the CP-violating phase from Ref.~\cite{Esteban:2016qun} have been used.}
\label{fig:UHE}
\end{figure}
%%%%%%%%%%%%%%%%%%%%%%%%%%%%%%%%%%%%%%%%%%%%%%%%%%%%%%%%%%%%%%%%%%%%%%%%%%%%%

In Fig.~\ref{fig:UHE}, we present the final results of the flavor ratios $\widehat{\Phi}^{\rm D}_\alpha/\Phi$ (with $\Phi \equiv \widehat{\Phi}^{\rm D}_e + \widehat{\Phi}^{\rm D}_\mu + \widehat{\Phi}^{\rm D}_\tau$) of high-energy cosmic neutrinos after the solar matter effects are taken into account. For comparison, the ratios in the absence of solar matter effects are also given and represented by dashed lines. In our numerical calculations, the exact transitional probabilities have been implemented without any approximations, and a constant matter density of $1.408~{\rm g}~{\rm cm}^{-3}$ with $Y^{}_e \approx 0.7$ has been adopted for the Sun. As we have observed before, as long as the condition $A \gg \Delta m^2_{31}$ is satisfied, the detailed density profile of the Sun is not important. In fact, we have demonstrated numerically that taking different values of the matter density does not alter our results much. The left two panels are for the cases of conventional $pp$ sources and the $\mu$-damped $pp$ sources, while the right two panels for the $p\gamma$ and $\mu$-damped $ p\gamma$ sources. A few interesting features of Fig.~\ref{fig:UHE} can be observed:

(1) In all four scenarios, one can see the oscillatory behaviors of the flavor ratios against neutrino energies in the presence of solar matter effects, while the flavor ratios in the standard case are independent of neutrino energies. In reality, the neutrino fluxes should be both energy- and flavor-dependent, but the dependence will be quite different from the oscillatory one under consideration. From Fig.~\ref{fig:UHE}, one can observe two different modes of oscillations, which are driven by the neutrino mass-squared differences $\Delta m^2_{21} = 7.5\times 10^{-5}~{\rm eV}^2$ and $\Delta m^2_{31} = 2.5\times 10^{-3}~{\rm eV}^2$ in vacuum, respectively. The amplitude of the low-frequency oscillations corresponding to $\Delta m^2_{21} = 7.5\times 10^{-5}~{\rm eV}^2$ is governed by the leading terms in Eq.~(\ref{eq:Vappro}), while that of the high-frequency oscillations by the higher-order terms $\mathcal{O}({\epsilon}^{1/2})$ $\sim 0.1$.

(2) The neutrino flavor ratios significantly deviate from the standard values at relatively low energies $E \sim 10~{\rm TeV}$, and become smaller for higher energies, gradually converging to the standard ratios for $E \gtrsim 10^3~{\rm TeV}$ in the case of $pp$ sources or for $E \gtrsim 10^4~{\rm TeV}$ in the case of $p\gamma$ sources. This different behavior can be understood by comparing the matter effects for neutrinos and those for antineutrinos. The key quantities are the interference terms $\widehat{I}^{}_{\alpha \beta}$ in the transitional probabilities. For neutrinos, we recast $\widehat{I}^{}_{\alpha \beta}$ into the following form
\begin{eqnarray}\label{eq:IabSunSeparated}
\widehat{I}^{}_{\alpha \beta} &=& +\sum^3_{i = 1} k^\alpha_i \sum_{j > k} 2 {\rm Re}\left( U^{}_{\beta j} U^*_{\beta k} A^{}_{ij} A^*_{ik} \right) \cos \left[ \Delta m^2_{jk} L^{}_{\rm SE}/(2E) \right] \; \nonumber \\
&& + \sum^3_{i = 1} k^\alpha_i \sum_{j > k} 2 {\rm Im}\left( U^{}_{\beta j} U^*_{\beta k} A^{}_{ij} A^*_{ik} \right) \sin \left[ \Delta m^2_{jk} L^{}_{\rm SE}/(2E) \right] \; .
%     (38)
\end{eqnarray}
For antineutrinos, we have to replace $U$ with $U^*$ and $A$ with $-A$, which introduces a minus sign to the imaginary part in the second line on the right-hand side of Eq.~(\ref{eq:IabSunSeparated}) and leads to
\begin{eqnarray}\label{eq:IabSunSeparatedAnti}
\widehat{I}^{}_{\overline{\alpha} \overline{\beta}} &=& +\sum^3_{i = 1} k^\alpha_i \sum_{j > k} 2 {\rm Re}\left( U^{}_{\beta j} U^*_{\beta k} A^{}_{ij} A^*_{ik} \right) \cos \left[\Delta m^2_{jk} L^{}_{\rm SE}/(2E) \right] \; \nonumber \\
&&- \sum^3_{i = 1} k^\alpha_i \sum_{j > k} 2 {\rm Im}\left( U^{}_{\beta j} U^*_{\beta k} A^{}_{ij} A^*_{ik} \right) \sin \left[ \Delta m^2_{jk} L^{}_{\rm SE}/(2E) \right] \;.
%     (39)
\end{eqnarray}
Therefore, for the $pp$ sources, the neutrino fluxes are equal to the antineutrino ones, so the terms in the second lines on the right-hand sides of Eqs.~(\ref{eq:IabSunSeparated}) and (\ref{eq:IabSunSeparatedAnti}) will cancel out. The remaining terms converge to the standard values of neutrino flavor ratios rapidly as the neutrino energy increases. However, for the $p\gamma$ sources, the neutrino and antineutrino fluxes are not exactly the same, retaining the terms that are more sensitive to neutrino energies and converge to the standard case slowly.

(3) The previous observations are applicable to the ratios $\widehat{\Phi}^{\rm D}_\mu/\Phi$ and $\widehat{\Phi}^{\rm D}_\tau/\Phi$ but not to $\widehat{\Phi}^{\rm D}_e/\Phi$, whose oscillatory behavior is almost universal for all four scenarios as shown in Fig.~\ref{fig:UHE}. The reason is that the terms ${\rm Im}\left( U^{}_{\beta j} U^*_{\beta k} A^{}_{ij} A^*_{ik} \right)$ with $\beta = e$ for all possible subscripts  in Eq.~(\ref{eq:IabSunSeparated}) and Eq.~(\ref{eq:IabSunSeparatedAnti}) vanish up to the order of $\mathcal{O}(\epsilon)\sim 0.01$, which can be verified by tracing the high-order terms in Eq.~(\ref{eq:Aijapprox}). However, this is not true for $\nu^{}_{\mu}$ and $\nu^{}_{\tau}$, as the terms of $\mathcal{O}(\epsilon^{1/2})\sim 0.1$ are retained.

Finally, it is worthwhile to mention that the attenuation of neutrino fluxes in the matter could also be very important due to the absorption of neutrinos, especially for the astrophysical object with a very large radius like the Sun. The attenuation length for the neutrinos of energies around $1~{\rm TeV}$ is almost comparable to the solar diameter \cite{Gandhi:1998ri}, thus would greatly weaken the visibility of the flavor conversions. Besides, the solar atmospheric neutrinos originated from cosmic rays interacting with the solar atmosphere could be a severe background of the astrophysical neutrinos traversing the Sun \cite{Arguelles:2017eao,Ng:2017aur,Edsjo:2017kjk,Masip:2017gvw}. For the neutrino telescopes like IceCube and KM3NeT, it is very difficult to observe the oscillatory behavior of the flavor composition caused by the solar matter effects. First of all, the Sun occupies only a small (i.e., $6\times 10^{-6}$) fraction of the total sky. The IceCube detector running for 5.7 years has observed 82 high-energy starting events~\cite{Aartsen:2017mau}, and the number of atmospheric neutrino background events is $40$. Since one needs 25 neutrino events from the Sun to achieve a statistical accuracy of $20\%$, the total event number registered in the IceCube detector should be around $4\times 10^6$ in assumption of an isotropic diffuse astrophysical neutrino flux. A realistic observation requires an upgrade of the fiducial volume of current detectors by a factor of $10^5$, which seems to be unfeasible even in the far future. A promising situation for the detection is that some strong transient point source happens to be behind the Sun, but such a possibility could also be very low.

\section{Summary}

In this work, we motivate the studies of the lunar matter effects on the solar neutrinos by asking if it is possible to observe the solar eclipses in the neutrino light. To answer this question, we set up the framework to investigate the regeneration effects of the initially decoherent neutrinos in the intermediate astrophysical objects. The original neutrino states are assumed to be decoherent superpositions of neutrino mass eigenstates, and the intermediate astrophysical object can be the Moon for solar neutrinos and the Sun for the high-energy cosmic neutrinos. The day-night asymmetry of high-energy solar neutrinos, which has been observed in the Super-Kamiokande experiment, serves as a typical example.

The essential idea is that the coherence among neutrino mass eigenstates can be regenerated by the intermediate astrophysical object. We have demonstrated that this kind of matter effects will be important if the regenerated coherence can be maintained along the whole way to the detector. Unfortunately, this is not the case for the lunar matter effects on solar neutrinos, since the distance between the Moon and the Sun is much longer than the relevant oscillation lengths. For this reason, the flavor conversions of solar neutrinos induced by the Moon are further reduced by a factor of $1.2\%$, compared to the day-night effect. On the other hand, the lunar matter effects on high-energy cosmic neutrinos of energies $E \gtrsim 10~{\rm TeV}$ are found to be as small as $0.1\%$ for a different reason, namely, the Moon's diameter and the distance between the Moon and the Earth are too short for the oscillations to develop. However, when the high-energy cosmic neutrinos traverse the Sun, the impact on their flavor ratios can be as large as $20\%$, this is mainly due to the long distance from the Sun to the Earth. This effect decreases with an increasing neutrino energy, and almost vanishes at the energies higher than $1~{\rm PeV}$. Although it is actually quite challenging to probe solar matter effects on high-energy cosmic neutrinos in the present neutrino telescopes, the novel matter effects  under discussions are interesting on their own.

\section*{Acknowledgment}

The authors would like to thank Zhi-zhong Xing and Yu-feng Li for helpful discussions. This work was supported in part by the National Natural Science Foundation of China under Grant No. 11775232, by the National Youth Thousand Talents Program and by the CAS Center for Excellence in Particle Physics (CCEPP).

\end{document}